\documentclass[aps,prd,preprint,floats,epsf,superscriptaddress,nofootinbib,amsmath]{revtex4}

\usepackage{graphicx}
\usepackage{amssymb,amsmath}
\usepackage{mathrsfs}
\usepackage{slashed}
\usepackage{indentfirst}
\usepackage{graphicx}
\usepackage{xcolor}
\usepackage{dsfont}
\usepackage{ulem}

\usepackage[colorlinks,
            linkcolor=black,
            anchorcolor=black,
            citecolor=black
            ]{hyperref}

\def\ie{{\it i.e.}}

\def\bsmm{$B_s \to \mu^+ \mu^-$ }

\def\BRbsmm{{\overline{\mathcal B}(B_s \to \mu^+ \mu^-)}}

\newcommand{\GeV}{{\, \rm GeV}}
\newcommand{\MeV}{\, \rm MeV}
\newcommand{\fb}{{\,\rm fb}}
\newcommand{\ps}{{\, \rm ps}}
\newcommand{\SM}{\text{SM}}
\newcommand{\NP}{\text{NP}}
\newcommand{\mcO}{\mathcal O}

\newcommand{\tblue}[1]{\textcolor{blue}{#1}}

\begin{document}
 \preprint{NCTS-PH/1704}

\bigskip

\title{Constraints and Implications on Higgs FCNC Couplings\\ from Precision Measurement of $B_s \to \mu^+ \mu^-$ Decay}

\author{Cheng-Wei Chiang}
\email[e-mail: ]{chengwei@phys.ntu.edu.tw}
\affiliation{Department of Physics, National Taiwan University, Taipei 10617, Taiwan}
\affiliation{Institute of Physics, Academia Sinica, Taipei 11529, Taiwan}
\affiliation{Physics Division, National Center for Theoretical Sciences, Hsinchu 30013, Taiwan}

\author{Xiao-Gang He}
\email[e-mail: ]{hexg@phys.ntu.edu.tw}
\affiliation{INPAC, Department of Physics and Astronomy, Shanghai Jiao Tong University, Shanghai 200240, China}
\affiliation{Department of Physics, National Taiwan University, Taipei 10617, Taiwan}
\affiliation{Physics Division, National Center for Theoretical Sciences, Hsinchu 30013, Taiwan}

\author{Fang Ye}
\email[e-mail: ]{fangye@ntu.edu.tw}
\affiliation{Department of Physics, National Taiwan University, Taipei 10617, Taiwan}

\author{Xing-Bo Yuan}
\email[e-mail: ]{xbyuan@cts.nthu.edu.tw}
\affiliation{Physics Division, National Center for Theoretical Sciences, Hsinchu 30013, Taiwan}

\date{\today}

\begin{abstract}
We study constraints and implications of the recent LHCb measurement of ${\cal B}(B_s \to \mu^+\mu^-)$ for tree-level Higgs-mediated flavor-changing neutral current (FCNC) interactions.  Combined with experimental data on $B_s$ mass difference $\Delta m_s$, the $h \to \mu \tau$, and the $h \to \tau^+\tau^-$ decay branching ratios from the LHC, we find that the Higgs FCNC couplings are severely constrained.  The allowed regions for $B_s \to \mu \tau$, $\tau\tau$ and $h \to sb$ decays are obtained.  Current data allow large CP violation in the $h \to \tau^+ \tau^-$ decay.  Consequences of the Cheng-Sher ansatz for the Higgs Yukawa couplings are discussed in some detail.  
\end{abstract}

\maketitle

\section{Introduction}

The Standard Model (SM) of particle physics~\cite{Glashow:1961tr,Weinberg:1967tq,Salam:1968rm,Englert:1964et,Higgs:1964pj,Guralnik:1964eu} has been working successfully to explain most phenomena observed in experiments.  It reached its summit when the 125-GeV Higgs boson was discovered~\cite{Aad:2012tfa,Chatrchyan:2012xdj} and its properties were later on shown to be 
in good agreement with SM expectations.  An on-going program in particle physics is to determine at high precision the Higgs couplings with other SM particles, as such studies could reveal whether there is an extended Higgs sector and give us more information about electroweak symmetry breaking.
If there is an extended Higgs sector, many observables in flavor physics that are sensitive to new physics (NP) can be affected.

In general, models with physics beyond the SM can lead to Higgs-mediated flavor-changing neutral currents (FCNC's)~\cite{Branco:2011iw}, which have severe constraints from flavor physics.  Even though such FCNC's can be avoided by imposing certain conditions for natural flavor conservation~\cite{Glashow:1976nt} or Yukawa alignment~\cite{Pich:2009sp}, it is better to leave it to experimental data to tell us whether the FCNC couplings are indeed negligibly small or sufficiently sizeable to have some intriguing phenomenological effects.

One channel that provides an excellent probe for the Higgs-mediated FCNC couplings is the rare \bsmm decay~\cite{Altmannshofer:2011gn,DeBruyn:2012wk,Fleischer:2012fy,Buras:2013uqa,Altmannshofer:2017wqy}.  This decay has a relatively simple structure in the SM, involving only a single vector current operator in the effective interaction Hamiltonian. Recently the  LHCb Collaboration has measured a branching ratio
$\BRbsmm_{\rm LHCb}=\bigl(3.0 \pm 0.6 ^{+0.3}_{-0.2}\bigr) \times 10^{-9}$ for the \bsmm decay~\cite{LHCb:2017}.  Combined with the previous CMS measurement $\BRbsmm_{\rm CMS} = \bigl(3.0_{-0.9}^{+1.0} \bigr) \times 10^{-9}$~\cite{Chatrchyan:2013bka}, one would obtain the average value 
\begin{eqnarray}\label{bsmm:avg}
\BRbsmm_{\rm avg}&=\bigl(3.0 \pm 0.5\bigr)\times 10^{-9} ~.
\end{eqnarray}
This value is in general agreement with the value predicted in the SM~\cite{Bobeth:2013uxa}, which, using currently known inputs, is
 \begin{eqnarray}\label{bsmm:SM}
 \BRbsmm_\SM=(3.44 \pm 0.19)\times 10^{-9}\;.
 \end{eqnarray}
Comparing the values above, one notices that the experimental central value is 
about $13\%$ lower than the SM one. NP effects may address such a discrepancy, though the error bars are still too large to call for such a solution.  Nevertheless one can use Eqs.~(\ref{bsmm:avg}) and (\ref{bsmm:SM}) to constrain possible NP contributions and study their implications.

If the 125-GeV Higgs boson $h$ has FCNC couplings to fermions, its mediation can produce scalar and/or pesudoscalar operators that contribute to the $B_s\to \mu^+\mu^-$ decay.  In the SM, such operators are generated only at loop level and further suppressed by the small muon Yukawa coupling.  
However, such interactions may be generated at tree level and do not suffer from chiral suppression in physics beyond the SM.  It is the primary purpose of this work to constrain such couplings using the recently measured $B_s\to \mu^+\mu^-$ decay along with others, and study the implications for other processes.

Another closely related and important constraint comes from the $B_s$ mass difference through the $B_s$-$\bar B_s$ mixing effect.  The updated SM prediction of $\Delta m_s$~\cite{Artuso:2015swg} and the most recent experimental measurement~\cite{Amhis:2016xyh} are, respectively,
\begin{eqnarray}\label{massdiff:Bs}
  \Delta m_s^\SM = \bigl(18.64 _{-2.27}^{+2.40}\bigr)~ \ps^{-1},\quad  \Delta m_s^{\rm exp} = (17.757 \pm 0.021)~ \ps^{-1}.
\end{eqnarray}
They provide a tight constraint on tree-level Higgs scalar and pseudoscalar couplings with the $s$ and $b$ quarks.

In the lepton sector, a hint of significant flavor-changing Higgs couplings,  ${\cal B}(h \to \mu \tau) = \bigl(0.84^{+0.39}_{-0.37}\bigr)\%$, was first reported by the CMS collaboration corresponding to an integrated luminosity of $19.7\,\fb^{-1}$~\cite{Khachatryan:2015kon}. However, recent measurements by the CMS and ATLAS,
\begin{align}\label{htomutau:exp}
{\cal B}(h \to \mu \tau)_{\rm CMS} <0.25\%~\text{\cite{CMS:2017onh}},
\qquad
{\cal B}(h \to \mu \tau)_{\rm ATLAS} < 1.43\% ~\text{\cite{Aad:2016blu}},
\end{align}
at the 95\%-CL have excluded the possibility of sizeable $\mu$-$\tau$ flavor-violating Higgs couplings indicated by 
the earlier CMS data~\cite{Khachatryan:2015kon}.

The existence of FCNC couplings of the Higgs boson to fermions occur in many extensions of the SM in the Higgs sector~\cite{Buras:2013rqa}, such as multi-Higgs doublet models~\cite{Branco:2011iw}.  
A simple example that can lead to tree level Higgs FCNC couplings with fermions is by introducing certain dimension-6 operators~\cite{Harnik:2012pb}: 
\begin{eqnarray}
{\phi^\dagger \phi\over\Lambda^2}\bar{\ell}_{Li} g^l_{ij}\phi e_{Rj}\;,\;\; 
{\phi^\dagger \phi\over\Lambda^2} \bar{Q}_{Li} g^d_{ij}\phi  D_{Rj}\;,\;\;{\phi^\dagger \phi\over\Lambda^2}\bar{Q}_{Li}g^u_{ij} \tilde \phi  U_{Rj}\;,
\end{eqnarray}
where $\Lambda$ denotes some new physics scale,
in addition to the usual dimension-4 Yukawa interactions
$\bar{\ell}_{Li} y^u_{ij} \phi \  e_{Rj}$,  $\bar{Q}_{Li} y^d_{ij}\phi  D_{Rj}$, and $\bar{Q}_{Li}y^u_{ij}\tilde \phi  U_{Rj}$.
Here $\ell_{Li}$ denote the left-handed leptons, $Q_{Li}$ the left-handed quarks, $e_{R_i}$ the right-handed charged leptons, $D_{R_i}$ the right-handed down-type quarks, $U_{R_i}$ the right-handed up-type quarks, $\phi$ the Higgs doublet,
and $\tilde \phi \equiv i\sigma_2 \phi^*$.

In the mass eigenbasis, Higgs FCNC interactions will be generated by the term $\delta Y^f = (v^2/2\Lambda^2) (S^\dagger_L g^f S_R)$ induced by the above-mentioned dimension-6 operators, where $S_L$ and $S_R$ denote respectively the bi-unitary transformation matrices for the left-handed and right-handed fermion fields to obtain the diagonal fermion mass matrix $\hat M$.  As a result, the Yukawa interaction Lagrangian in the mass eigenbasis is given by
\begin{eqnarray} \label{eq:Lagrangian:FCNC}
&&{\cal L}_{h\bar f f }
\equiv - \frac{1}{\sqrt{2}} \bar f (Y^f + i \gamma_5\bar Y^f) f h ~,
\end{eqnarray}
where $Y^f = \sqrt{2}\hat M^f/v + (\delta Y^f + \delta Y^{f \dagger})$ and 
$\bar Y^f = -i(\delta Y^f - \delta Y^{f \dagger})$ are in general non-diagonal and $v = 246$~GeV is the vacuum expectation value of the SM Higgs field.  Hence, they can induce Higgs-mediated FCNC processes at tree level.

In this work, we  
make use of the combined result of the \bsmm branching ratio, the $B_s$ mass difference~(\ref{massdiff:Bs}), and the $h \to \tau\tau$~\cite{Khachatryan:2016vau} and $h \to \mu \tau$~\cite{Khachatryan:2015kon,Aad:2016blu} decay rates to constrain the involved Higgs couplings.  From the constrained parameter space, we can make predictions for the $B_s \to \mu^\pm \tau^\mp$ and $\tau^+ \tau^-$ as well as the $h \to sb$ decays without invoking additional assumptions.

Generically elements in the Yukawa matrices $Y^f$ and $\bar{Y}^f$ are independent of each other. In order to increase the predictive power, one often employs some texture for the Yukawa couplings, such as the Cheng-Sher ansatz~\cite{Cheng:1987rs}, so that one can also compute the rates for more related processes.  We will 
take the Cheng-Sher ansatz as a working assumption to put it to a test in the face of the coupling constraints extracted from the above-mentioned data.  

The structure of this paper is as follows.  In Section~\ref{sec:Bsmumu}, we discuss how the tree-level Higgs-mediated FCNC interactions affect the $B_s \to \mu^+\mu^-$ decay, the $B_s$-$\bar B_s$ mixing, and the $h \to \mu \tau$ and $\tau\tau$ decays.  In Section~\ref{sec:predictions}, we present a detailed numerical analysis to obtain the allowed parameter space for the FCNC couplings.  In Section~\ref{sec:h-tau-mu}, we first study implications for the $h\to \mu \tau$ and $B_s \to \mu\tau$, $\tau\tau$ decays without invoking any additional  assumptions.  We then estimate more related observables by taking the Cheng-Sher ansatz.  We draw conclusions in Section~\ref{sec:summary}.

\section{Theoretical Framework \label{sec:Bsmumu}}

In this section, we discuss how the Higgs Yukawa couplings given in Eq.~(\ref{eq:Lagrangian:FCNC}) affect the processes of interest to us; namely, the \bsmm decay, the $B_s$ mass difference, and the $h \to \mu\tau$ and $sb$ decays.

\subsection{The \bsmm decay}

With the Higgs exchanges introduced in the previous section and the SM contribution, the effective Hamiltonian responsible for the ${\bar B}_s \to \mu^+ \mu^-$ decay is given by~\cite{Buchalla:1995vs}
\begin{eqnarray}\label{eq:Hamiltonian}
  \mathcal H_{\rm eff}
  = -\frac{G_F}{\sqrt 2} \frac{\alpha_{em}}{\pi s_W^2} V_{tb}V_{ts}^*
  \bigl(C_A\mathcal O_A + C_S\mathcal O_S + C_P\mathcal O_P 
  + C'_S\mathcal O'_S + C'_P\mathcal O'_P \bigr)+h.c.,
\end{eqnarray}
where $\alpha_{em}$ is the fine structure constant, and $s_W^2 \equiv \sin^2\theta_W$ with $\theta_W$ being the weak mixing angle.  $V_{ij}$ denote the Cabibbo-Kobayashi-Maskawa (CKM) matrix elements. The operators $O_i^{(\prime)}$ are defined as
\begin{align}\label{eq:operator}
\mcO_A &=\bigl(\bar q \gamma_\mu P_L b\bigr)\bigl(\bar\mu \gamma^\mu\gamma_5 \mu\bigr)\;,
&
\mcO_S &= m_b \bigl(\bar q P_R b\bigr)\bigl(\bar \mu \mu\bigr)\;,
&
\mcO_P &= m_b \bigl(\bar q P_R b\bigr)\bigl(\bar\mu \gamma_5 \mu\bigr)\;,\nonumber\\
&&
\mcO_S^\prime&= m_b \bigl(\bar q P_L b\bigr)\bigl(\bar \mu \mu\bigr)\;,
&
\mcO_P^\prime&= m_b \bigl(\bar q P_L b\bigr)\bigl(\bar\mu \gamma_5 \mu\bigr)\;,
\end{align}
where the $b$ quark mass $m_b$ is included in the definition of ${\cal O}^{(\prime)}_{S,P}$ so that their Wilson coefficients are renormalization group invariant~\cite{Altmannshofer:2011gn}.

In the framework we are working with, the Wilson coefficient $C_A$ contains only the SM contribution, and its explicit expression up to the NLO QCD corrections can be found in Refs.~\cite{Buchalla:1993bv,Misiak:1999yg,Buchalla:1998ba}.  Recently, corrections at the NLO EW~\cite{Bobeth:2013tba} and NNLO QCD~\cite{Hermann:2013kca} have been completed, with the numerical value approximated by~\cite{Bobeth:2013uxa}
 \begin{eqnarray}
 &&C_A^{\SM}(\mu_b) = -0.4690 \left(\frac{m_t^{\rm P}}{173.1~\mbox{GeV}}\right)^{1.53} \left(\frac{\alpha_s(m_Z)}{0.1184}\right)^{-0.09}\;,
  \end{eqnarray}
where $m_t^{\rm P}$ denotes the top-quark pole mass.  In the SM, the Wilson coefficients $C_S^\SM$ and $C_P^\SM$ can be induced by the Higgs-penguin diagrams but are highly suppressed.  Their expressions can be found in Refs.~\cite{Li:2014fea,Cheng:2015yfu}.  As a very good approximation, we can safely take $C_S^\SM=C_S^{\prime \SM}=C_P^\SM=C_P^{\prime \SM}=0$.

With the Higgs-mediated FCNC interactions in the effective Lagrangian, Eq.~(\ref{eq:Lagrangian:FCNC}), the scalar and pseudoscalar Wilson coefficients 
\begin{align}
C_S^{\NP} &= \kappa (Y_{sb} + i \bar Y_{sb})  Y_{\mu\mu}\;,& C_P^{\NP} &=i\kappa (Y_{sb} + i \bar Y_{sb})\bar Y_{\mu\mu}\;, \nonumber\\
C_S^{\prime \NP} &= \kappa (Y_{sb} - i \bar Y_{sb}) Y_{\mu\mu}\;,& C_P^{\prime \NP} &= i\kappa (Y_{sb} - i\bar Y_{sb}) \bar Y_{\mu\mu}\;,
\end{align}
where the common factor 
\begin{align}
\kappa=\frac{\pi^2}{2G_F^2}\frac{1}{V_{tb}V_{ts}^*}\frac{1}{m_b m_h^2m_W^2}\;.
\end{align}

For the effective Hamiltonian Eq.~\eqref{eq:Hamiltonian}, the branching ratio of \bsmm reads~\cite{Li:2014fea,Cheng:2015yfu}
\begin{eqnarray}
  \mathcal B(B_s \to \mu^+\mu^-) = 
  \frac{\tau_{B_s}G_F^4 m_W^4}{8\pi^5} |V_{tb}V_{tq}^*|^2 f_{B_s}^2 m_{B_s} m_\mu^2
  \sqrt{1-\frac{4 m_\mu^2}{m_{B_s}^2}} \bigl(|P|^2+|S|^2\bigr)\;,
\end{eqnarray}
where $m_{B_s}$, $\tau_{B_s}$ and $f_{B_s}$ denotes the mass, lifetime and decay constant of the $B_s$ meson, respectively. The amplitudes $P$ and $S$ are defined as
\begin{eqnarray}
  &&
  P\equiv
  C_A+\frac{m_{B_s}^2}{2 m_\mu}\left(\frac{m_b}{m_b+m_s}\right)(C_P-C_P^\prime)\;,\nonumber\\
  &&
  S \equiv 
  \sqrt{1-\frac{4m_\mu^2}{m_{B_s}^2}}\frac{m_{B_s}^2}{2 m_\mu}
  \left(\frac{m_b}{m_b+m_s}\right)(C_S-C_S^\prime)\;.
\end{eqnarray}

Note that the NP scalar operators (\ie, the $\bar{Y}_{sb}Y_{\mu \mu}$ term) contribute to the branching ratio incoherently and always increase the latter, while the NP pseudoscalar operators (\ie, the $\bar{Y}_{sb}\bar{Y}_{\mu\mu}$ term) have interference with the SM amplitude and the resulting effects may be constructive or destructive, depending on the sign of $\bar{Y}_{sb}\bar{Y}_{\mu\mu}$. Given that the experimental value of the branching ratio is lower than that predicted by the SM, we expect  the $\bar{Y}_{sb}\bar{Y}_{\mu\mu}$ parameter to play the role of reducing the \bsmm theoretical value to the experimental level.

Due to the $B_s$-$\bar B_s$ oscillations, the measured branching ratio of \bsmm should be the time-integrated one~\cite{DeBruyn:2012wk}:
\begin{align}
  \BRbsmm=\left(\frac{1+\mathcal A_{\Delta\Gamma}y_s}{1-y_s^2}\right)\mathcal
  B(B_s\to \mu^+\mu^-)\;,
\end{align}
where~\cite{Buras:2013uqa}
\begin{align}\label{eq:ys}
  y_s = \frac{\Gamma_s^{\rm L}-\Gamma_s^{\rm H}}{\Gamma_s^{\rm L} + \Gamma_s^{\rm H}}=\frac{\Delta\Gamma_s}{2\Gamma_s}
  \quad \text{and} \quad
   \mathcal A_{\Delta\Gamma}=\frac{|P|^2\cos {(2\varphi_P-\phi_s^\NP)}-|S|^2\cos{( 2\varphi_S-\phi_s^\NP)}}{|P|^2+|S|^2}\;,
\end{align}
$\Gamma_s^{\rm L}$ and $\Gamma_s^{\rm H}$ denote respectively the decay widths of the light and heavy $B_s$ mass eigenstates, and $\varphi_P$ and $\varphi_S$ are the phases associated with $P$ and $S$, respectively. The CP phase $\phi_s^\NP$ comes from $B_s$-$\bar B_s$ mixing and will be defined in eq.~\eqref{eq:dltm}. In the SM, $\mathcal A_{\Delta\Gamma}^\SM=1$.

\subsection{The mass difference $\Delta m_s$ \label{sec:dMs}}

If $Y_{sb}$ and/or $\bar Y_{sb}$ are non-zero, contributions to $B_s$-$\bar B_s$ mixing can be induced.  Therefore, one must make sure that the current measurement of mass difference $\Delta m_s$ is respected. 
In the SM, $B_s$-$\bar B_s$ mixing occurs mainly via the box diagrams involving the exchange of $W^\pm$ bosons and top quarks.  The mass difference between the two mass eigenstates $B_s^H$ and $B_s^L$ can be obtained from the $\Delta B=2$ effective Hamiltonian~\cite{Buras:2001ra}  
\begin{eqnarray}\label{eq:Hamiltonian:mixing}
\mathcal H_{\rm eff}^{\Delta B=2} 
= \frac{G_F^2}{16\pi^2}m_W^2 (V_{tb}V_{ts}^*)^2\sum_i  C_i \mathcal O_i + {\rm h.c.} ~,
\end{eqnarray}
where the operators relevant to our study are
\begin{align}
  \mathcal O_1^{\rm VLL}&=(\bar s^\alpha\gamma_\mu P_L b^\alpha)(\bar s ^\beta \gamma^\mu P_L b^\beta)\;,&
  \mathcal O_1^{\rm SLL}&=(\bar s^\alpha P_L b^\alpha)(\bar s^\beta P_L b^\beta)\;,\nonumber\\
  \mathcal O_2^{\rm LR}&=(\bar s^\alpha P_L b^\alpha)(\bar s^\beta P_R b^\beta)\;,&
  \mathcal O_1^{\rm SRR}&=(\bar s^\alpha P_R b^\alpha)(\bar s^\beta P_R b^\beta)\;,
\end{align}
with $\alpha$ and $\beta$ color indices.  The SM contributes only the $\mathcal O^{\rm VLL}_1$ operator, with the corresponding Wilson coefficient at the LO given by~\cite{Buras:1998raa}
  \begin{eqnarray}
    C_1^{\rm VLL,SM}(\mu_W) \approx 9.84 \left(\frac{m_t}{170\GeV} \right)^{1.52}\;,
  \end{eqnarray}
whose analytical expression can be found in Ref.~\cite{Buchalla:1995vs}.

With the effective Lagrangian in Eq.~\eqref{eq:Lagrangian:FCNC}, the tree-level Higgs exchange results in 
\begin{align}\label{eq:WC:mixing}
    C_1^{\rm SLL, \NP} &= -\frac{1}{2}\tilde  \kappa (Y_{sb}-i\bar Y_{sb})^2\;,&
    C_2^{\rm LR, \NP} &=-\tilde \kappa (Y_{sb}^2+\bar Y_{sb}^2)\;,\nonumber\\
    C_1^{\rm SRR, \NP} &= -\frac{1}{2}\tilde \kappa (Y_{sb} + i\bar Y_{sb})^2\;,&
    \tilde \kappa  &=\frac{8\pi^2}{G_F^2} \frac{1}{m_h^2m_W^2}\frac{1}{(V_{tb}V_{ts}^*)^2}\;.
\end{align}

The contribution from $\mathcal H_{\rm eff}^{\Delta B=2}$ to the transition matrix element of $B_s - \bar B_s$ mixing is given by~\cite{Buras:2001ra},
\begin{align}
  M_{12}^s
= \langle B_s | \mathcal H_{\rm eff}^{\Delta B=2} | \bar B_s \rangle
= \frac{G_F^2}{16\pi^2} m_W^2 (V_{tb}V_{ts}^*)^2  \sum C_i \langle B_s \left\lvert \mathcal O_i \right\rvert \bar B_s \rangle\,,
\end{align}
where recent lattice calculations of the hadronic matrix elements $\langle \mathcal O_i \rangle$ can be found in Refs.~\cite{Carrasco:2013zta,Bazavov:2016nty}. Then the mass difference and CP violation phase read
\begin{align}\label{eq:dltm}
  \Delta m_s = 2 |M_{12}^s|\,, \qquad \text{and} \qquad \phi_s= \arg M_{12}^s\,.
\end{align}
 In the case of complex Yukawa couplings, $\phi_s$ can derivate from the SM prediction, i.e., $\phi_s=\phi_s^\SM+\phi_s^\NP$. Nonzero $\phi_s^\NP$ can affect the CP violation in the $B_s \to J/\psi \phi$ decay~\cite{Artuso:2015swg}, as well as $\mathcal A_{\Delta\Gamma}$ in the \bsmm decay as in eq.~\eqref{eq:ys}. We note that $\Delta m_s$ depends only on $Y_{sb}^2$ and $\bar Y_{sb}^2$, but not $Y_{sb}\bar Y_{sb}$.  In addition, we follow Ref.~\cite{Buras:2001ra} to perform renormalization group evolution of the NP operators $\mathcal O_1^{\rm SLL}$, $\mathcal O_1^{\rm SRR}$ and $\mathcal O_2^{\rm LR}$.  It is found that including RG effects of the NP operators enhances the NP contributions by about a factor of 2.

\subsection{The $h \to f_1 f_2$ decays \label{sec:hff}}

The partial width of the Higgs boson decaying to a pair of fermions in the Born approximation is given by
\begin{align}
\Gamma(h \to f_1 f_2)
= S N_c \frac{m_h}{8\pi} \left(
|Y_{f_1f_2}|^2 + |{\bar Y}_{f_1f_2}|^2
\right) ~,
\label{eq:Gamhsb}
\end{align}
where $S = 1 ~(1/2)$ when $f_1$ and $f_2$ are of different (same) flavors and $N_c$ denotes the number of colors for the fermions.

With the pseudoscalar Yukawa couplings also included in our analysis, one can consider the possibility of observing CP violation in $h \to \tau^+ \tau^-$ through the operator
${\cal O}_\pi = \vec{p}_\tau\cdot(\vec{p}_{\pi^+} \times \vec{ p}_{\pi^-})$.  Here 
$\vec p_{\pi^+}$ and $\vec p_{\pi^-}$ are respectively the 3-momenta of $\pi^+$ and $\pi^-$ from the 
$ \tau^+\rightarrow \pi^+\bar \nu_{\tau}$ and $\tau^- \rightarrow \pi^-\nu_{\tau}$ decay, and $\vec p_\tau$ is the momentum of the $\tau^-$ from $h\to \tau^+ \tau^-$ decay. Letting $N_+$ and $N_-$ be events with 
${\cal O}_\pi >0$ and ${\cal O}_\pi <0$, respectively, one can define a CP violating observable
\begin{eqnarray}
A_\pi =\frac{N_+ -N_- }{N_+ +N_- }\approx
\frac{ \pi}{ 4} \frac{ (Y_{\tau\tau} \bar Y_{\tau\tau} )}{ Y^2_{\tau\tau} + \bar Y^2_{\tau\tau}  }\;,
\label{piasym}
\end{eqnarray}
which can be measured experimentally~\cite{hmg}.

\section{NUMERICAL ANALYSIS \label{sec:predictions}}

\begin{table}[t]
  \centering
  \begin{tabular}{lllll l}
    \hline\hline
    Input & Value & Unit & Ref. 
  \\\hline
  $\alpha_s^{(5)}(m_Z)$ & $0.1181\pm 0.0011$ & & \cite{Olive:2016xmw}
  \\
  $1/\alpha_{\rm em}^{(5)}(m_Z)$& $127.944 \pm 0.014$&& \cite{Olive:2016xmw}
  \\
  $m_t^{\rm P}$ & $173.21 \pm 0.51 \pm 0.71$ & GeV & \cite{Olive:2016xmw}
  \\
  \hline
  $|V_{cb}|$ (semi-leptonic) & $41.00 \pm 0.33 \pm 0.74$ &$10^{-3}$& \cite{Charles:2004jd}
  \\
  $|V_{ub}|$ (semi-leptonic) & $3.98 \pm 0.08 \pm 0.22$ & $10^{-3}$& \cite{Charles:2004jd}
  \\
  $|V_{us}|f_+^{K\to\pi}(0)$ &  $0.2165 \pm 0.0004$ & & \cite{Charles:2004jd}
  \\
  $\gamma$ &  $72.1_{-5.8}^{+5.4}$ & $[^\circ]$ & \cite{Charles:2004jd}
  \\
  $f_+^{K\to \pi}(0)$ & $0.9681 \pm 0.0014 \pm 0.0022$ & & \cite{Charles:2004jd}
  \\
  \hline
  $f_{B_s}$ & $228.4\pm 3.7$ & MeV & \cite{Aoki:2016frl}
  \\
  $f_{B_s}\sqrt{\hat B}$  & $270\pm 16$ & MeV & \cite{Aoki:2016frl}
  \\
  $1/\Gamma_s^{\rm H}$  & $1.609 \pm 0.010$ & ps & \cite{Amhis:2016xyh}
  \\
  $\Delta\Gamma_s/\Gamma_s$ & $0.129\pm 0.009$ & & \cite{Amhis:2016xyh}
  \\
  \hline\hline
  \end{tabular}
  \caption{Inputs for $B_s \to \mu^+ \mu^-$ and $B_s$-$\bar B_s$ mixing.}
  \label{tab:input}
\end{table}

With the theoretical formalism discussed in the previous sections and the input parameters given in Table~\ref{tab:input}, we can compare relevant SM predictions with the recent experimental measurements to see if any NP is allowed.

At present, the theoretical uncertainties for $B_s\to \mu^+\mu^-$ and $\Delta m_s$ mainly arise from the decay constant $f_{B_s}$ and the CKM matrix element $|V_{cb}|$. As is well known, there is a long-standing tension between the inclusive and exclusive determinations of $|V_{cb}|$ and $|V_{ub}|$~\cite{Olive:2016xmw}.  We find that the branching ratio obtained from the exclusive $|V_{cb}|$ and $|V_{ub}|$ are about $10\%$ smaller than the one from the inclusive values, mainly due to the difference in $|V_{cb}|$.  Here we adopt the recent average given by the CKMfitter group~\cite{Charles:2004jd}.  For the lifetime, both $\Gamma_s^{\rm H}$ and $\Delta\Gamma_s/\Gamma_s$ are used.  The SM prediction then depends only on $\Gamma_s^{\rm H}$. Finally, compared to the SM prediction of $(3.65 \pm 0.23) \times 10^{-9}$ previously given in Ref.~\cite{Bobeth:2013uxa}, our theoretical uncertainty is smaller mainly due to more precise values of $f_{B_s}$ and $\Gamma_s^{\rm H}$.

For the $B_s$-$\bar B_s$ mixing, the SM prediction of $\Delta m_s$ in Ref.~\cite{Artuso:2015swg} is updated with the input parameters in Table~\ref{tab:input}, and reads in comparison with the most recent experimental measurement~(\ref{massdiff:Bs}). 
Note that the SM central value is larger than the experimental one.
Hence we expect the NP amplitude to interfere with the SM amplitude destructively. We will see later that this results in an upper bound of the Yukawa couplings $|\bar{Y}_{sb}|$ and $|Y_{sb}|$.

In the following, we carry out numerical analysis for constraints on the Yukawa couplings from \bsmm and $B_s$-$\bar B_s$ mixing.  The allowed parameter space of these Yukawa couplings from each of the observables is obtained by requiring that the difference between the theoretical prediction and experimental measurement be less than twice the error bar (i.e. 95\% confidence level (CL).), calculated by adding the theoretical and experimental errors in quadrature.

\begin{figure}[t]
    \centering
    \includegraphics[width=0.45\textwidth]{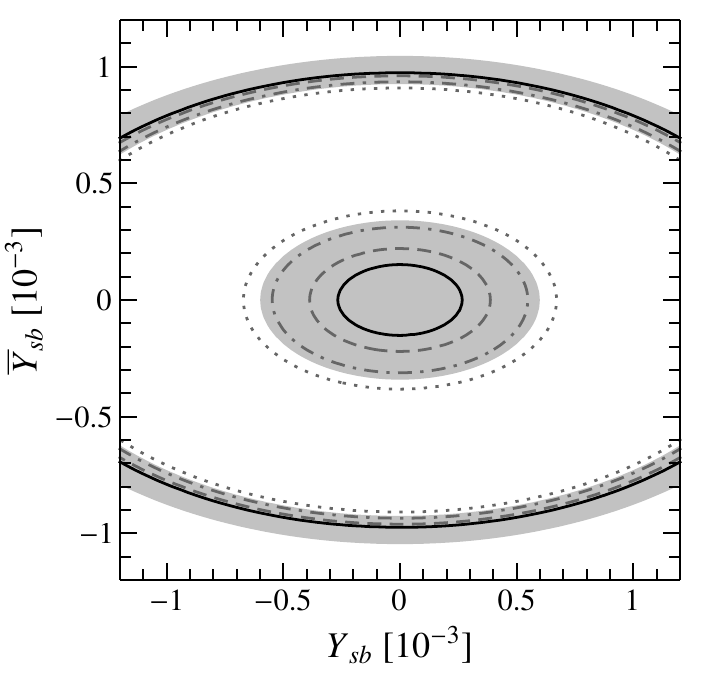}
    \caption{Allowed parameter space in the $(Y_{sb}, \bar Y_{sb})$ plane as constrained by the $B_s$-$\bar B_s$ mixing.  The black solid curve and the shaded region correspond respectively to the central value and the 95\%-CL region of the measured $\Delta m_s$.  The dashed, dot-dashed, and dotted contours correspond to $\Delta m_s/\Delta m_s^\SM=$ 0.9, 0.8 and 0.7, respectively.}
    \label{fig:Bs_mixing}
  \end{figure}

Fig.~\ref{fig:Bs_mixing} shows the constraints in the $(Y_{sb},\,\bar Y_{sb})$ plane from the $B_s$-$\bar B_s$ mixing, assuming the two Yukawa couplings to be real.  As mentioned earlier, we find two shaded regions that agree with the experimental measurement at 95\% CL. Near the origin in the parameter space, the Higgs-mediated FCNC effects are mostly destructive with the SM contributions.
In this region, the pseudoscalar coupling $\bar Y_{sb}$ has the bound
\begin{align}
\label{eq:YsbBound}
|\bar Y_{sb}| \alt 3.4 \times 10^{-4}\,.
\end{align}
The outer elliptical band corresponds to the case where the Higgs-mediated FCNC interactions dominate over the SM contribution, thus flipping the sign of $M_{12}^s$. The corresponding bound on $|\bar Y_{sb}|$ is $0.9 \times 10^{-3} \alt |\bar Y_{sb}| \alt 1.1 \times 10^{-3}$.  We do not pursue this possibility in the following analysis.

\begin{figure}[t]
    \centering
    \includegraphics[width=0.45\textwidth]{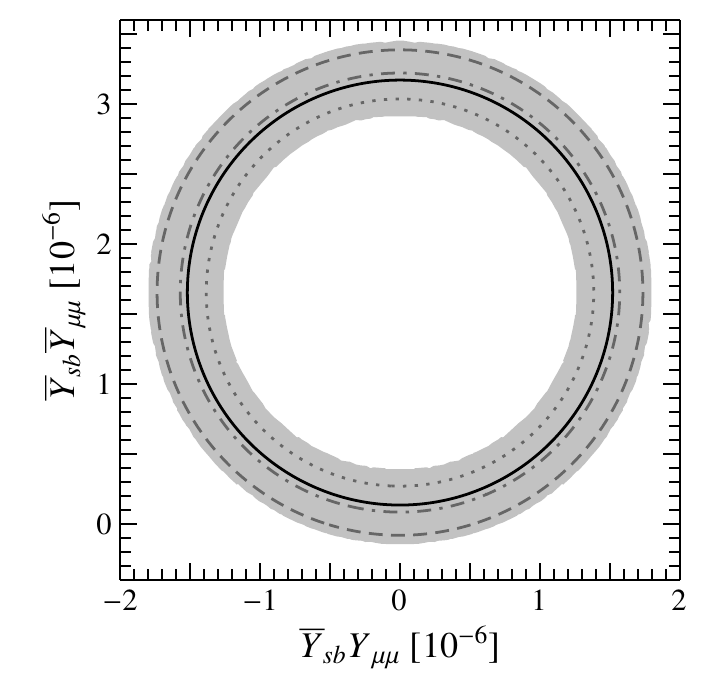}
    \includegraphics[width=0.45\textwidth]{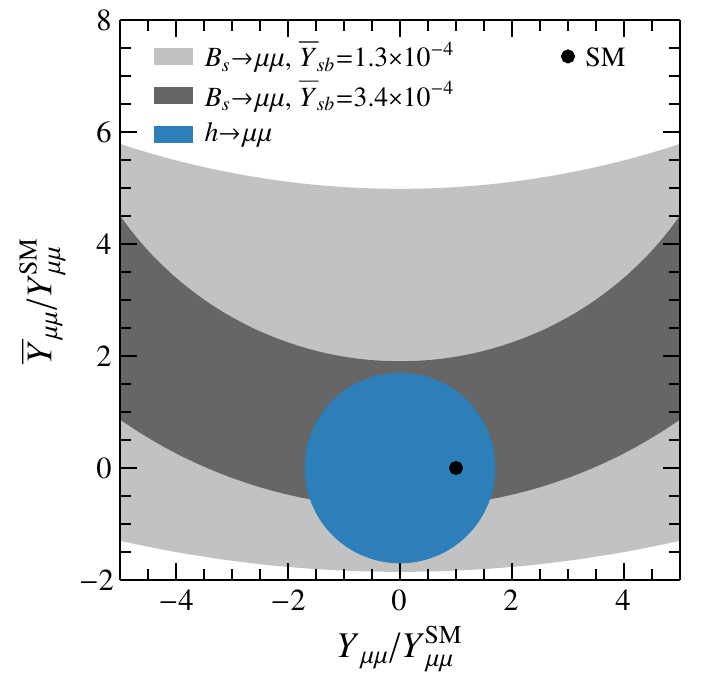}
    \caption{Left: Allowed parameter space in the $(\bar Y_{sb}Y_{\mu\mu},\,\bar Y_{sb}\bar Y_{\mu\mu})$ plane as constrained by the \bsmm decay.  The black solid curve and the shaded region correspond respectively to the central value and the allowed region at 95\% CL.  The dashed, dot-dashed and dotted contours correspond to $\BRbsmm/\BRbsmm_\SM=$ 1.1, 0.9 and 0.7, respectively. Right: Allowed region for $(Y_{\mu\mu}/ Y_{\mu\mu}^\SM,\, \bar Y_{\mu\mu}/Y_{\mu\mu}^\SM)$ obtained for the choices of $\bar Y_{sb}=3.4\times 10^{-4}$ (dark gray) and $\bar Y_{sb}=1.3\times 10^{-4}$ (light gray), as well as from the direct measurement of $h \to \mu^+ \mu^-$ at the LHC (blue). The black point indicates the SM Yukawa couplings.}
    \label{fig:general}
  \end{figure}

With the contributions from the Higgs FCNC Lagrangian in Eq.~\eqref{eq:Lagrangian:FCNC}, the branching ratio of \bsmm depends on two parameters: $\bar Y_{sb} Y_{\mu\mu}$ and $\bar Y_{sb} \bar Y_{\mu\mu}$. The combined CMS and LHCb measurement of $\BRbsmm$ at 95\% CL implies the following bounds:
\begin{align}
{0.66} \alt \bigl\lvert 5.6\times 10^5\,\bar Y_{sb}Y_{\mu\mu}\bigr\rvert^2 + \bigl\lvert 1-6.0\times 10^5\,\bar Y_{sb}\bar Y_{\mu\mu}\bigr\rvert^2 \alt {1.26}\;.
\label{range-mumu}
\end{align}
For illustration purposes, we have taken the Yukawa couplings to be real, and plot the allowed region for $\bar Y_{sb}Y_{\mu\mu}$ and $\bar Y_{sb}\bar Y_{\mu\mu}$ in the left plot of Fig.~\ref{fig:general}.
As discussed in the previous section, the NP pseudoscalar operator $\mcO_P^\NP$ ({\it i.e.}, the $\bar Y_{sb} \bar Y_{\mu\mu}$ contribution) has either constructive or destructive interference with the SM amplitude, while the NP scalar operator $\mcO_S^\NP$ ({\it i.e.}, the $\bar{Y}_{sb}Y_{\mu\mu}$ contribution) always enhances the branching ratio. 
Therefore, in the region of small $\bar Y_{sb}Y_{\mu\mu}$ and $\bar Y_{sb} \bar Y_{\mu\mu}$, 
the branching ratio is much more sensitive to the parameter $\bar Y_{sb}Y_{\mu\mu}$ than to $\bar Y_{sb} \bar Y_{\mu\mu}$. It is also noted that the current experimental central value $\BRbsmm_{\text{avg}}/\BRbsmm_\SM \approx 0.87$.

Taking the largest value $\bar Y_{sb}= 3.4 \times 10^{-4}$,  allowed by $B_s -\bar B_s$ mixing in the central region in Fig.~\ref{fig:Bs_mixing}, and a relative small Yukawa coupling $\bar Y_{sb}= 1.3 \times 10^{-4}$ as two explicit examples, we then obtain the right plot of Fig.~\ref{fig:general} that shows a closer view of the muon Yukawa couplings in the vicinity of their SM values.  Apparently, the region allowed by the former (depicted in dark gray) and that by the latter (depicted in light gray) are parts of two annular rings, respectively.  To further limit the allowed parameter space, we use the Higgs signal strength of the muon channel, $\mu_{\mu\mu}<2.8$ at 95\% CL recently reported by ATLAS~\cite{Aaboud:2017ojs} from a combination of the 7\,TeV, 8\,TeV and 13\,TeV ATLAS data.  This is given by the blue circular area.  As a consequence, the pseudoscalar muon Yukawa coupling is restricted to $|\bar Y_{\mu\mu}| \alt 1.7\, Y_{\mu\mu}^\SM$.

\begin{figure}[t]
  \centering
  \includegraphics[width=0.7\textwidth]{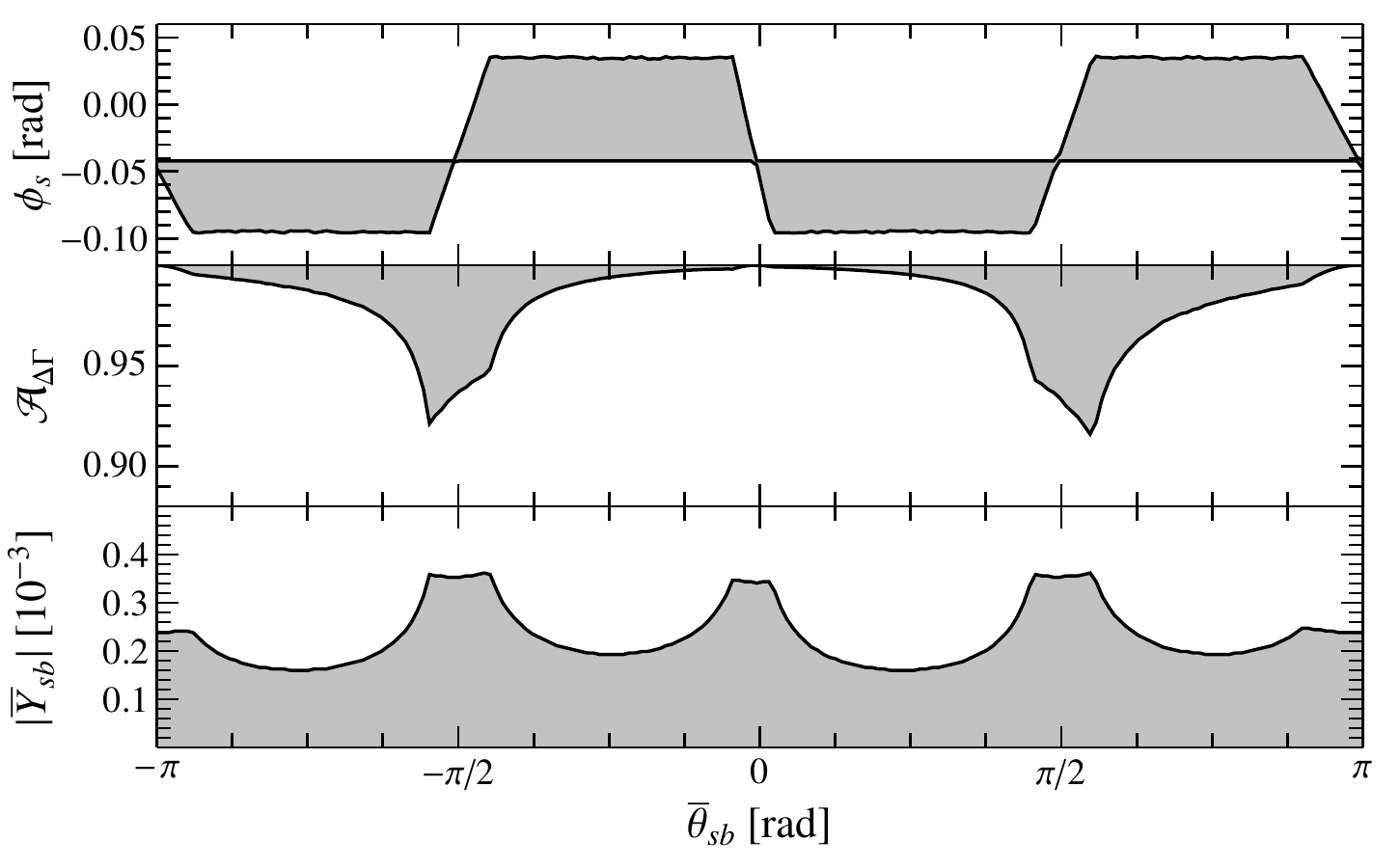}
  \caption{Allowed parameter space of $(\bar\theta_{sb},|\bar Y_{sb}|)$, as constrained by the \bsmm and $B_s$-$\bar B_s$ mixing at 95\% CL under the assumption of $(Y_{sb},Y_{\mu\mu},\bar Y_{\mu\mu})=(0,Y_{\mu\mu}^\SM,Y_{\mu\mu}^\SM)$.  The corresponding allowed regions for $\mathcal A_{\Delta\Gamma}$ and $\phi_s$ are also given.}
  \label{fig:phase}
\end{figure}

For the $B_s$-$\bar B_s$ mixing, when allowing the Yukawa couplings to be complex, the 95\% CL bound changes to
\begin{align}
\label{eq:complexYsb}
0.76 \alt 
\bigl\lvert 1 - \left(0.7\, Y_{sb}^2 + 2.1 \, \bar Y_{sb}^2\right)\times 10^6 \bigr\rvert 
\alt 1.29\;.
\end{align}
In this case, several new effects show up.  The phase $\phi_s\equiv\arg(M_{12}^s)$ for $B_s$-$\bar B_s$ mixing will acquire a non-vanishing NP piece, and this will affect the parameter $\mathcal A_{\Delta\Gamma}$. We have carried out a numerical analysis considering experimental bounds from these quantities. As an illustration, we take $(Y_{sb},Y_{\mu\mu},\bar Y_{\mu\mu}) = (0,Y_{\mu\mu}^\SM,Y_{\mu\mu}^\SM)$ and obtain the bounds on the phase $\bar \theta_{sb}$ and the magnitude of $\bar Y_{sb}$ from the \bsmm decay and $B_s$-$\bar B_s$ mixing. Fig.~\ref{fig:phase} shows the allowed parameter space for $(\bar\theta_{sb}, |\bar Y_{sb}|)$ and the corresponding regions of $\mathcal A_{\Delta\Gamma}$ and $\phi_s$.  As can be seen in Eq.~\eqref{eq:ys}, the Higgs FCNC effects on $\mathcal A_{\Delta\Gamma}$ become significant for $\bar\theta_{sb}\approx$ $\pm \pi/2$ under the current assumption of $Y_{sb} = 0$.  As the SM contribution has an almost null phase in $M_{12}^s$, $\phi_s$ has a significant modification when $\bar\theta_{sb}\approx\pm \pi/4$ or $\pm 3\pi/4$.  Since we have included the constraints from the CP phase $\phi_s^{c \bar c s} = -0.03 \pm 0.033$ radian~\cite{Amhis:2016xyh}, the regions near $\bar \theta_{sb}\approx \pm\pi/4,\pm 3\pi/4$ are more strongly constrained.


There are also constraints from the $h \to \mu \tau$ data from the LHC.  Very recently, a new search based on a dataset of $35.9$~fb$^{-1}$ at the CMS results in an upper bound $\mathcal B ( h \to \mu \tau) < 0.25\%$~\cite{CMS:2017onh}, which excludes the previous hint of sizeable $\mu$-$\tau$ flavor-violating Higgs couplings.  Here the complex $Y_{\mu\tau}$ and $\bar Y_{\mu\tau}$ can contribute to the $h \to \mu\tau$ decay at tree level, and one has from the new CMS data that~\cite{CMS:2017onh}
\begin{align}
\label{eq:complexYmt}
\sqrt{|Y_{\mu\tau}|^2 + |\bar Y_{\mu\tau}|^2 } < 1.43 \times 10^{-3}
\end{align}
at $95\%$ CL.  This imposes a very stringent restraint on the NP effects, to be discussed in the next section.

\section{Discussions \label{sec:h-tau-mu}}

In the previous section, we have shown that the precision measurements of the \bsmm decay and $\Delta m_s$ have tightly restricted the allowed ranges of some tree-level Higgs FCNC interactions.  With the input of $h \to \mu \tau$ decay width, we have also obtained restraints in a couple of lepton FCNC Yukawa couplings.  It is remarkable that flavor physics has now become a precision test ground for the study of Higgs properties.  We now discuss the implications of the above-mentioned constraints in other rare decay processes.

\subsection{The $h \to sb$, $B_s \to \tau\tau$, and $B_s \to \mu\tau$ decays}

In the SM, the Higgs total decay width $\Gamma_h^{\rm SM} \simeq 4.1$~MeV.  This can be modified if the $h \to sb$ and $\mu\tau$ decay considered in this work contribute significantly.  Using the constraint Eq.~\eqref{eq:complexYsb} obtained for the generally complex Yukawa couplings in Section~\ref{sec:dMs}, we have
\begin{align}
\Gamma(h \to sb) < 0.043 ~{\MeV}
~~\mbox{or}~~
{\cal B}(h \to sb) < 1.05\%
\end{align}
at 95\% CL.  Note that here we only consider the scenario where the SM contribution dominates in the estimate of $\Delta m_s$.  With such a small decay rate and only one $b$ quark for tagging, the channel is expected to be very difficult to measure at the LHC.

The $B_s \to \tau^+\tau^-$ decay rate calculation is similar to that of the \bsmm decay.  
Using the experimental $h\to \tau\tau$ data and constraints on  the generally complex $Y_{sb}$ and $\bar Y_{sb}$ from $B_s-\bar{B}_s$ mixing, we find
\begin{align}
0.6 ~(0.5) 
< \frac{\mathcal B(B_s \to \tau^+\tau^-)}{\mathcal B(B_s \to \tau^+ \tau^-)_\SM}
< 1.5 ~(1.7)
\end{align} 
at $1\sigma$ level (95\% CL).

In the SM, the $B_s \to \mu\tau$ decay is suppressed because its leading-order process occurs at the one-loop level and the neutrino mass (difference) is extremely small.  However, with the FCNC couplings assumed in Eq.~\eqref{eq:Lagrangian:FCNC}, this decay process happens at tree level through the mediation of the Higgs boson.  Again, by scanning the allowed parameter space given in Eqs.~\eqref{eq:complexYsb} and \eqref{eq:complexYmt}, we find that 
${\cal B}(B_s \to \mu\tau)$ can be as large as $0.8 ~(1.8) \times 10^{-8}$
at $1\sigma$ level (95\% CL).  The 95\%-CL upper limit is about one order of magnitude larger than the currently measured \bsmm decay branching ratio.

\subsection{Leptonic decays of $B_s$ and $h$ with the Cheng-Sher ansatz}

As mentioned earlier, the flavor-conserving and flavor-changing components of the scalar and pseudoscalar Yukawa couplings $Y$ and $\bar Y$ are generally independent.  To improve the predictive power, one can assume some specific relations among the couplings.  One popular scienario is the Cheng-Sher ansatz~\cite{Cheng:1987rs}. One can apply the Cheng-Sher ansatz to the quark and lepton sectors separately.  As an illustration, here we will work with only applying the Cheng-Sher ansatz to the dimension-6 operators involving charged leptons to see how some predictions can be made.  In this case, the charged lepton Yukawa couplings take the following form
\begin{align}
\label{eq:Cheng-Sher}
    Y_{ij} = \delta_{ij}\frac{\sqrt{2}m_i}{v} + \xi_\ell \frac{\sqrt{2m_im_j}}{v}
    \qquad \text{and} \qquad 
    \bar Y_{ij} = \bar\xi_\ell \frac{\sqrt{2m_im_j}}{v} ~,
 \end{align}
where $\xi_\ell$ and $\bar \xi_\ell$ vanish in the SM limit.

\begin{figure}[t]
  \centering
  \includegraphics[width=0.45\textwidth]{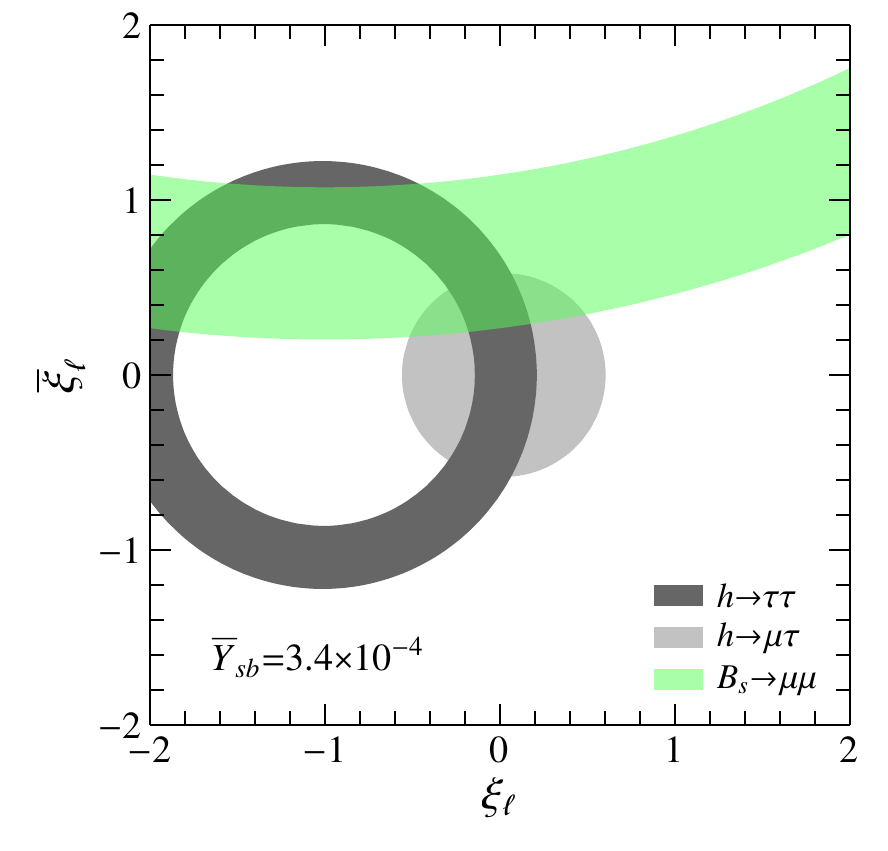}
  \quad
  \includegraphics[width=0.45\textwidth]{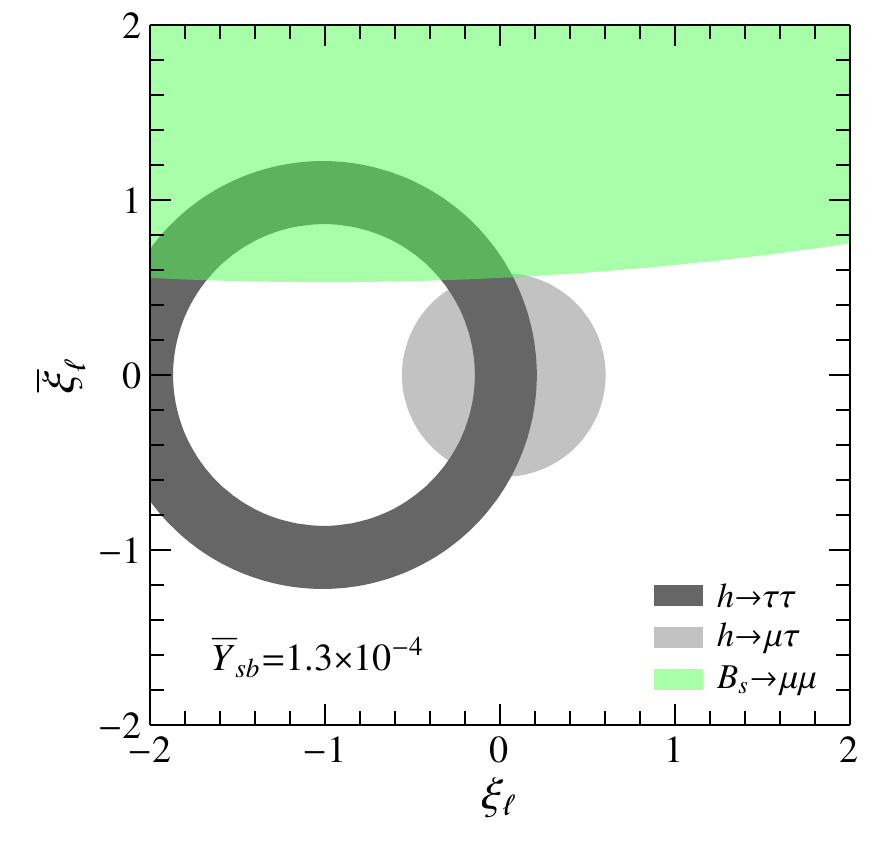}
  \caption{Combined constraints on $(\xi_\ell,\bar\xi_\ell)$ under the Cheng-Sher ansatz. The dark and light gray regions are respectively the parameter space allowed by the $h \to \tau\tau$ measurement from the combined \sout{Run-I} \tblue{LHC} data and the CMS $h \to \mu \tau$ data at 95\% CL. In the case of $\bar Y_{sb}= 3.4 \times 10^{-4}$ (left plot) and $\bar Y_{sb}= 1.3 \times 10^{-4}$ (right plot), the parameter space satisfying $75\% < \mathcal B(B_s \to \mu^+\mu^-)/\mathcal B(B_s \to \mu^+ \mu^-)_\SM < 95\%$ is shown by the green region.}
  \label{fig:combined}
\end{figure}

In the following, we will apply this ansatz and take into account the new upper bound ${\cal B}(h \to \mu \tau)_{\rm CMS} < 0.25\%$~\cite{CMS:2017onh} and the signal strength of the $h \to \tau\tau$ channel, $\mu_{\tau\tau} = 1.11_{-0.22}^{+0.24}$~\cite{Khachatryan:2016vau} measured at the Run I LHC and $\mu_{\tau\tau}=1.06_{-0.24}^{+0.25}$ recently measured by CMS at 13\,TeV with a dataset of $35.9\,\fb^{-1}$~\cite{CMS:2017wyg}.  We will also use Eq.~\eqref{eq:Gamhsb} to predict the flavor-changing $h \to sb$ decay rate.

With the Cheng-Sher ansatz in the lepton sector, the constraints on $(Y_{\mu\tau},\bar Y_{\mu\tau})$ from the CMS data can be converted to the constraints on $(\xi_\ell, \bar\xi_\ell)$, as shown in Fig.~\ref{fig:combined}, where the subscript $\ell$ refers to the charged leptons.  In this figure, the parameter regions allowed by the $h \to \tau\tau$ measurement from the combined LHC data and the new CMS bound on $\mathcal B(h \to \mu \tau)$ are respectively given by the dark gray ring and light gray circular area, both at 95\% CL.  Furthermore, as discussed in the previous section, the largest allowed $|\bar Y_{sb}|$ is about $3.4\times 10^{-4}$.  We take $\bar Y_{sb} = 3.4 \times 10^{-4}$ (left plot) and $1.3 \times 10^{-4}$ (right plot) as two benchmark values, and find the region in the $(\xi_\ell,\bar\xi_\ell)$ plane that reduces $\mathcal B(B_s \to \mu^+\mu^-)_\SM$ by $5\%$ to $25\%$, to be in better agreement with the current data.  It is shown that, unless for a very small $\bar Y_{sb}$, the Higgs FCNC couplings under the Cheng-Sher ansatz can simultaneously be consistent with the LHC Higgs measurements while suppressing the $B_s \to \mu^+ \mu^-$ branching ratio by $\sim 15\%$.  By varying $\bar Y_{sb}$ until there is no overlap between the region allowed by the $h \to \mu\tau$ and $\tau\tau$ data and the region for $5\%$ to $25\%$ reduction from $\mathcal B(B_s \to \mu^+\mu^-)_\SM$, one can obtain a lower bound on $|\bar Y_{sb}|$, as can be seen by comparing the left and right plots of Fig.~\ref{fig:combined}.  This exercise shows that if $\mathcal B(B_s \to \mu^+\mu^-)$ can be better determined and seen to be significantly lower than the SM prediction, a lower bound on the pseudoscalar FCNC Yukawa coupling $|\bar Y_{sb}|$ can be obtained, complementary to the upper bound from $\Delta m_s$ given in Eq.~\eqref{eq:YsbBound}.  The experimental data on $h\to \tau\tau$ play an important role in constraining the central region in Fig.~\ref{fig:combined}.

We note in passing that for the overlapped region between the green region and the light gray region in Fig.~\ref{fig:combined} and assuming that the up-type quarks have only the SM Yukawa couplings, the measured signal strengths of different Higgs decay channels are modified because the changes in their branching ratios.  The predictions under the Cheng-Sher ansatz are consistent with the current measurements.

\begin{figure}[t]
  \centering
  \includegraphics[width=0.45\textwidth]{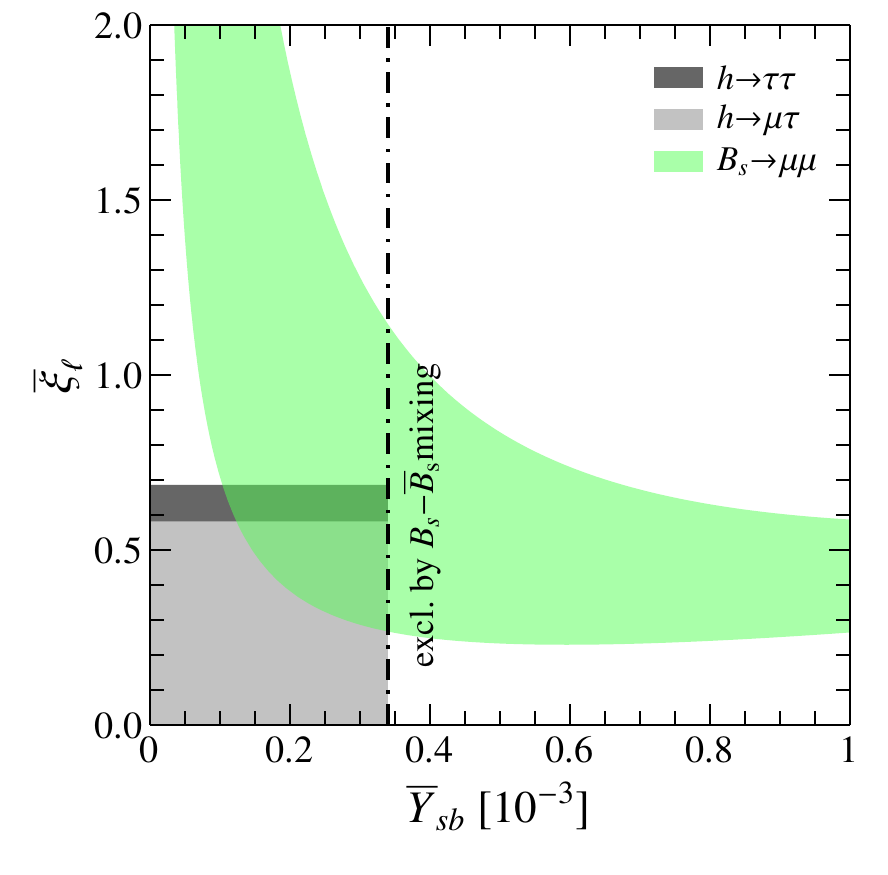}
  \caption{Combined constraints on $(\bar Y_{sb}, \bar\xi_\ell)$ under the Cheng-Sher ansatz in the case where scalar Yukawa couplings are purely SM-like.  The dot-dashed line denotes the upper bound from $\Delta m_s$. The plotting style is the same as in Fig.~\ref{fig:combined}.}\label{fig:combined_special}
\end{figure}

As an illustration to show the power of various experimental measurements, we consider the scenario where the scalar Yukawa couplings are SM-like ({\it e.g.}, $Y_{sb} = \xi_\ell = 0$) and the Cheng-Sher ansatz is applied only to the pseudoscalar Yukawa couplings of the charged leptons, {\it i.e.}, $\bar\xi_\ell\not= 0$.  Again, we take into account the measurements of $\Delta m_s$, \bsmm, $h \to \mu \tau$ and $h \to \tau\tau$ and show the combined constraints in the $(\bar Y_{sb}, \bar\xi_\ell)$ plane in Fig.~\ref{fig:combined_special}.  The light green region is plotted under the presumption that ${\cal B}(B_s \to \mu^+ \mu^-)$ will be measured with a higher precision and determined to fall between $75\%$ and $95\%$ of its SM expectation.  The region to the left of the dot-dashed line is ruled out by $\Delta m_s$ at $95\%$ CL.  The light gray region simultaneously satisfies the Higgs signal strength of the $\tau\tau$ channel and the new CMS upper bound on ${\cal B}(h \to \mu\tau)$ at $95\%$ CL.  The overlapped region (the greenish wedge at the upper right corner of the light gray area) shows nontrivial upper and lower bounds on the pseudoscalar parameters: $\bar\xi_\ell \in (0.27,\,0.58)$ and $\bar Y_{sb} \in (1.3,\,3.4) \times 10^{-4}$.  Such a scenario can be probed by future LHC and Belle-II experiments.

As alluded to in Section~\ref{sec:hff}, here we make a brief comment on the possibility of observing CP violation in $h \to \tau \bar \tau$ through the operator ${\cal O}_\pi = \vec{p}_\tau\cdot(\vec{p}_{\pi^+} \times \vec{ p}_{\pi^-})$.  
Taking $\bar Y_{sb} = 3.4 \times 10^{-4}$, as in the left plot of Fig.~\ref{fig:combined}, one can infer using the Cheng-Sher ansatz for $Y_{\tau\tau}$ and $\bar Y_{\tau\tau}$ that the absolute value of $A_\pi$, defined in Eq.~(\ref{piasym}), can be almost as large as the maximally allowed value of $\pi/8$.  This can be tested at a Higgs factory.

If one also applies the Cheng-Sher ansatz to the down-type quarks, rough estimates of the Higgs FCNC contributions to $\Delta m^{\rm NP}_{K}$ and $\Delta m^{\rm NP}_{B_d}$ can be made once $\Delta m^\NP_{B_s}$ is known, using 
\begin{align}
\begin{split}
\Delta m^{\rm NP}_K \approx 
\frac{R_K}{R_{B_s}} \frac{f^2_K m_K}{f_{B_s}^2 m_{B_s}}
\frac{m_d}{m_b} \Delta m^{\rm NP}_s ~,
\\
\Delta m^{\rm NP}_d \approx 
\frac{R_{B_d}}{R_{B_s}} \frac{f^2_{B_d} m_{B_d}}{f_{B_s}^2 m_{B_s}}
\frac{m_d}{m_s} \Delta m^{\rm NP}_s ~,
\end{split}
\end{align}
where $R_K / R_{B_s} \simeq 12.6$ and $R_{B_d} / R_{B_s} \simeq 1$~\cite{Buras:2001ra}, and the last fractions in both expressions come from the ansatz.  Assuming $\Delta 
m^{\rm NP}_s$ to be about 10\% of the experimental value, we find that the contributions to $\Delta m^{\rm NP}_K$ is about 20\% of its experimental value, but with opposite sign for real Yukawa couplings. With complex Yukawa couplings, the contribution from the imaginary part will add to the SM predicted value and become closer to the experimental value. Since there is a large uncertainty caused by long distance contribution for $\Delta m_K$~\cite{Antonelli:1996qd,Bertolini:1997ir,Buras:2010pza,Buras:2014maa}, it is possible that when adding all contributions together the correct value will be produced and the Cheng-Sher ansatz is valid here.  The contributions to $\Delta m_d^{\rm NP}$ is also about 10\%.
Therefore within the region allowed by the $B_s$-$\bar B_s$ mixing, the $B_d$-$\bar B_d$ is predicted to be consistent with the data.  As a consequence, the $B_d \to \mu^+\mu^-$ decay branching ratio will also be about the same order as that in the SM, which is smaller than the current experimental bound of $3.4\times 10^{-10}$ at 95\% CL~\cite{LHCb:2017}.  One also predicts $\mathcal B(B_d \to \mu\tau )/\mathcal B(B_s \to \mu \tau)\approx m_d/m_s$ resulting in $\mathcal B(B_d \to \mu\tau ) <  1.5 \times 10^{-9}$. This is much smaller than current experimental bound of $2.2\times 10^{-5}$.

\section{Summary \label{sec:summary}}

Motivated by the recent precision determination of the \bsmm decay branching ratio, we consider its constraints on tree-level flavor-changing Yukawa couplings with the 125-GeV Higgs boson, as defined in Eq.~\eqref{eq:Lagrangian:FCNC}.  To gain more definite information, we also take into account the $B_s$ mass difference $\Delta m_s$, the $h \to \mu \tau$ decay branching ratio determined by the CMS Collaboration, and the signal strength of the $h \to \tau^+\tau^-$ channel from the combined LHC data.

In what follows, we summarize the constraints on flavor-changing couplings obtained in this work, assuming that they are generally complex.  From $\BRbsmm$ alone, we obtain
\begin{align*}
{0.66} \alt \bigl\lvert 5.6\times 10^5\,\bar Y_{sb}Y_{\mu\mu}\bigr\rvert^2 + \bigl\lvert 1-6.0\times 10^5\,\bar Y_{sb}\bar Y_{\mu\mu}\bigr\rvert^2 \alt {1.26} ~.
\end{align*}
From $\Delta m_s$, we have
\begin{align*}
0.76 \alt 
\bigl\lvert 1 - \left( 0.7 \, Y_{sb}^2 + 2.1 \, \bar Y_{sb}^2\right)\times 10^6 \bigr\rvert 
\alt 1.29 ~.
\end{align*}
Combining with the constraints from the $h \to \mu\tau$ branching ratio bound measured by the CMS Collaboration and the $h \to \tau\tau$ signal strength from the \sout{LHC Run-I} \tblue{combined LHC} data, we have made predictions for the branching ratios of $B_s \to \mu \tau$, $\tau\tau$ and $h \to sb$ decays.  In particular, ${\cal B}(B_s \to \mu \tau)$ can be as large as $3.1 \times 10^{-8}$ at 95\% CL.  This may be quite challenging for the LHCb and future Belle-II experiments to measure.

Finally, we use the above-mentioned constraints obtained from data to test the Cheng-Sher ansatz.
We have shown that if the \bsmm branching ratio is found to deviate significantly from the SM expectation in the future, the combined analysis with the $h \to \tau\tau$ and $\mu\tau$ data can give us a lower bound on the pseudoscalar Yukawa coupling $\bar Y_{sb}$, provided that the $B_s$ mass difference is still dominated by the SM contribution.  As an example, the parameter $\bar\xi_\ell$ is found to fall within the $(0.27,0.58)$ region when the scalar Yukawa couplings are assumed to be SM-like.  We have also made a brief comment on the possibility of observing CP violation in the $h \to \tau^+ \tau^-$ decay.

\section*{Acknowledgments}
CWC was supported in part by the Ministry of Science and Technology (MOST) of ROC (Grant No.~MOST~104-2628-M-002-014-MY4).  XGH was supported in part by MOE Academic Excellent Program (Grant No.~105R891505) and MOST of ROC (Grant No.~MOST~104-2112-M-002-015-MY3), and in part by NSFC of PRC (Grant No.~11575111). This work was also supported by Key Laboratory for Particle Physics, Astrophysics and Cosmology, Ministry of Education, and Shanghai Key Laboratory for Particle Physics and Cosmology (SKLPPC) (Grant No.~11DZ2260700).

\end{document}